\documentclass[iop]{emulateapj}

\usepackage{xspace}

\newcommand\cxo{\textit{Chandra}\xspace}
\newcommand\xmm{\textit{XMM-Newton}\xspace}
\newcommand\zp{$\zeta$~Pup\xspace}

\newcommand\apec{{\sc apec}\xspace}
\newcommand\windtabs{\textit{windtabs}\xspace}
\newcommand\tbabs{\textit{tbabs}\xspace}
\newcommand\wabs{\textit{wabs}\xspace}

\begin{document}

\title{Modeling broadband X-ray absorption of massive star winds}

\author{Maurice A. Leutenegger\altaffilmark{1,}\altaffilmark{2}, 
  David H. Cohen\altaffilmark{3}, 
  Janos Zsarg\'{o}\altaffilmark{4,}\altaffilmark{5},
  Erin M. Martell\altaffilmark{6,}\altaffilmark{3},
  James P. MacArthur\altaffilmark{3}, 
  Stanley P. Owocki\altaffilmark{7}, 
  Marc Gagn\'{e}\altaffilmark{8}, 
  D. John Hillier\altaffilmark{5}} 
\altaffiltext{1}{Laboratory for High Energy Astrophysics, Code 662,
  NASA/Goddard Space Flight Center, Greenbelt, MD 20771, USA;
  \email{Maurice.A.Leutenegger@nasa.gov}}
\altaffiltext{2}{NASA Postdoctoral Fellow} 
\altaffiltext{3}{Department of Physics and Astronomy, Swarthmore
  College, Swarthmore, PA 19081, USA}
\altaffiltext{4}{Instituto Politecnico Nacional, Escuela Superior de Fisica y Matematicas, C.P. 07738, Mexico, D. F., Mexico}
\altaffiltext{5}{Department of Physics and Astronomy, University of
  Pittsburgh, 3941 O'Hara Street, Pittsburgh, PA 15260, USA}
\altaffiltext{6}{Department of Astronomy, University of Chicago,
  Chicago, IL 60637, USA}
\altaffiltext{7}{Bartol Research Institute, University of Delaware,
  Newark, DE 19716, USA} 
\altaffiltext{8}{Department of Geology and Astronomy, West Chester
  University of Pennsylvania, West Chester, PA 19383, USA}

\shorttitle{X-ray absorption of massive star winds}
\shortauthors{Leutenegger et al.}

\begin{abstract}

  We present a method for computing the net transmission of X-rays
  emitted by shock-heated plasma distributed throughout a partially
  optically thick stellar wind from a massive star.  We find the
  transmission by an exact integration of the formal solution,
  assuming that the emitting plasma and absorbing plasma are mixed at
  a constant mass ratio above some minimum radius, below which there
  is assumed to be no emission. This model is more realistic than
  either the slab absorption associated with a corona at the base of
  the wind or the exospheric approximation that assumes that all
  observed X-rays are emitted without attenuation from above the
  radius of optical depth unity.  Our model is implemented in XSPEC as
  a pre-calculated table that can be coupled to a user-defined table
  of the wavelength dependent wind opacity.  We provide a default wind
  opacity model that is more representative of real wind opacities
  than the commonly used neutral interstellar medium (ISM)
  tabulation. Preliminary modeling of \cxo\ grating data indicates
  that the X-ray hardness trend of OB stars with spectral subtype can
  largely be understood as a wind absorption effect.

\end{abstract}


\section{Introduction}
\label{sec:intro}

The absorption of soft X-rays by the powerful, radiation-driven winds
of OB stars has long been recognized as a significant effect both on
the X-rays observed from these stars and on the physical conditions in
their winds.  The soft X-ray emission observed in OB stars by {\it
  Einstein} and {\it ROSAT} implied only modest wind attenuation of
the X-rays, and thus ruled out significant coronal emission as a
source of the ubiquitous X-ray emission seen in these massive stars
\citep{CO79, CS83, Hel93, 1994ApJ...437..351M}.  For this and other
reasons, the wind-shock paradigm for the production of X-rays in OB
stars has become generally accepted \citep[e.g.,][]{OCR88, PKPBH94,
  FKPPP97, FPP97, Kel01}, although many aspects are still poorly
understood.

Using X-ray observations to constrain X-ray production mechanisms
requires proper account of the \emph{absorption} of \emph{distributed}
sources of X-ray emission. The amount and wavelength dependence of the
wind absorption can be used as a diagnostic of the
location/distribution of the shock-heated plasma and of the wind
mass-loss rate, especially in terms of its effect on individual line
profile shapes \citep[hereafter OC01]{OC01}. Even simply deriving an
intrinsic X-ray luminosity for energy budget considerations requires
correctly modeling the significant attenuation of the emitting X-rays,
especially in the dense winds of O supergiants \citep{Hel93,
  1999ApJ...520..833O}.

Because the emitting plasma is spatially distributed throughout the
absorbing wind, simple prescriptions for the attenuation can be
inaccurate. Specifically, the commonly used slab model of absorption,
appropriate for an intervening interstellar medium (ISM) cloud where
all of the emission originates beyond the absorbing medium, has
transmission with an exponential dependence on slab optical depth, and
strongly overestimates the amount of attenuation as the wind becomes
optically thick. Furthermore, because hydrogen and helium are ionized,
models for neutral gas significantly overestimate wind
opacity. Unfortunately, due to the lack of easily available tools for
incorporating appropriate radiative transfer and opacities, inadequate
models, such as those intended for neutral interstellar medium
absorption, are routinely applied to account for wind
attenuation. This is in spite of the fact that a number of previous
works have recognized the necessity for and applied appropriate wind
absorption prescriptions \citep{Hel93, PHL01, OC01, OFH06}.

We have developed a method for implementing an exact solution to a
realistic model of the radiation transport that can be easily combined
with a pre-calculated opacity table to find the wavelength-dependent
emergent X-ray flux from a stellar wind. We also provide a reasonable
default opacity model that can be used for most OB star winds. Our
analysis tool, which we name \windtabs, for \textit{wind}
\textit{t}able \textit{abs}orption, can be combined with an
independent plasma emission model, such as the Astrophysical Plasma
Emission Code (\apec) \citep{SBLR01} that is widely used in fitting
stellar X-ray spectra.  This can be used to realistically model the
low-resolution CCD spectra that are produced in large quantities by
surveys of clusters and OB associations with \cxo\ and \xmm
\citep[e.g.,][]{2006MNRAS.372..661S, 2008ApJ...675..464W}.  It can
also be used to model grating spectra in detail, and provides a means
of disentangling the wind absorption effects from the emission
temperature effects that both appear to contribute to the recently
discovered trend in the morphology of OB star spectra observed at high
resolution with the \cxo gratings \citep{WNW09}.

\section{Radiation transport model}
\label{sec:radtran}

In this section, we derive an expression for the fraction of X-rays
transmitted from a massive star wind as a function of the
wavelength-dependent opacity, the mass-loss rate, and the wind
velocity law. We make assumptions similar to those made in
\citet{OC01}: we model the wind as a spherically symmetric
two-component medium, where a small fraction of the wind is heated to
X-ray emitting temperatures ($T_{\mathrm X} \sim 1-10\, \mathrm{MK}$),
while the bulk of the wind is composed of relatively cool material
($T_\mathrm{wind} / T_\mathrm{eff} \sim 0.5-1$) that can absorb the
X-rays via the bound-free opacity of the moderately ionized metals. We
assume that the X-ray emission turns on at some radius $R_0 > R_*$. We
also assume that the temperature distribution of the X-ray emitting
plasma is the same over the entire emitting volume.

The assumptions we make regarding the distribution of the X-ray
emitting plasma are rooted in the available observational evidence, as
well as in the results of extensive theoretical simulations. Detailed
studies of emission line profiles as well as constraints from
forbidden-to-intercombination line ratios in He-like ions support a
picture in which X-ray emission starts at $R_0 \sim 1.5 R_*$ for all
observable ions \citep{Cel06, LPKC06, Cohen2010}, with a roughly
constant filling factor above the onset radius. Numerical simulations
have typically found onset radii for strong shocks that are comparable
to this \citep{OCR88}.  \citet{FPP97} performed simulations of winds
seeded with base perturbations. They have found that clump--clump
collisions are important, and that the resulting X-ray emission is
distributed primarily below 10 stellar radii. \citet{RO02} have also
found X-ray emission with onset radii of order 1.5 stellar radii and
extending out to large radii.  However, the results for the onset
radius are also sensitive to the treatment of the scattered radiation
field \citep{1996ApJ...462..894O, 2009AIPC.1171..173O}.

The observed X-ray luminosity as a function of wavelength is given by
\begin{equation}
  L_\lambda = 4\pi\, \int\, dV\, \eta_\lambda(r)\, e^{-\tau(r, \mu, \lambda)}\, ,
  \label{eq:lx}
\end{equation}
where $\eta_\lambda(r)$ is the X-ray emissivity, and $\tau(r, \mu,
\lambda)$ is the continuum optical depth of the dominant cool
component along a ray from the emitting volume element to the
observer.

The optical depth can be derived as in \citet{OC01} for a smooth wind.
It is given by the integral
\begin{equation}
  \tau (p,z,\lambda) = \int_{z}^{\infty}\, dz^\prime\, \kappa(\lambda)\, \rho
  (r^\prime)\, .
\end{equation}
Here $p$ and $z$ are ray coordinates, with impact parameter $p =
\sqrt{1 - \mu^2}\, r$ and distance along the ray $z = \mu\, r$, where
$\mu$ is the direction cosine to the observer at local radius $r$.
$\kappa(\lambda)$ is the atomic opacity of the wind, and $\rho(r)$ is
the density of the wind.  Using the continuity equation, $\rho(r) =
\dot{M} / 4\pi r^2 v(r)$, where $\dot{M}$ is the stellar mass loss
rate, and defining the {\it characteristic wind optical depth},
\begin{equation}
  \tau_* \equiv \frac{\kappa(\lambda)\, \dot{M}}{4\pi R_* v_\infty}\, ,
  \label{eq:taustardefinition}
\end{equation}
we can write
\begin{equation}
  \tau (p,z) = \tau_*\,  t (p,z)\, ,
  \label{eq:tautotaustar}
\end{equation}
where
\begin{equation}
  t (p,z) \equiv \int _{z}^{\infty}\, \frac{R_* dz^\prime}{r^{\prime2}
    w(r^{\prime})}
  \label{eqn:tpz}
\end{equation}
is a dimensionless integral that depends purely on the ray geometry.
Here, $R_*$ is the stellar radius, $v_\infty$ is the wind terminal
velocity, and $w(r) \equiv v(r) / v_\infty$ is the scaled wind
velocity. Note that we have assumed that $\kappa(\lambda)$ is constant
throughout the wind; we further discuss this assumption in
Section~\ref{sec:opacity}.  We take the velocity to follow a beta law: $v =
v_\infty (1 - R_* / r)^\beta$. We also take $\beta = 1$ in this paper
as a good approximation for many O star winds; however, evaluation for
general values of $\beta$ is not difficult.

The emissivity is assumed to scale with density squared, as in
\citet{OC01}. We ignore the Doppler shift of the emitted X-rays and write
\begin{equation}
  \eta_\lambda (r > R_0) = \eta_{\lambda,0} \frac{\rho^2(r)}{\rho^2_0}\, .
  \label{eq:eta}
\end{equation}
This expression gives the correct radial dependence for the total line
strength, but discards information about the profile shape. This is a
justified approximation in calculating the broadband X-ray
transmission of the wind.  Here we assume that X-ray emission begins
at a minimum radius $R_0$, with $\rho_0 \equiv \rho (R_0)$ and
$\eta_{\lambda,0} \equiv \eta_{\lambda} (R_0)$. In this paper, we will
assume that the X-ray filling factor is constant with radius; it is
trivial to add a power-law radial dependence, as in OC01. We also
assume that X-ray emissivity follows the same radial distribution at
all observable wavelengths.

\begin{figure}
  \includegraphics[width=3.3in]{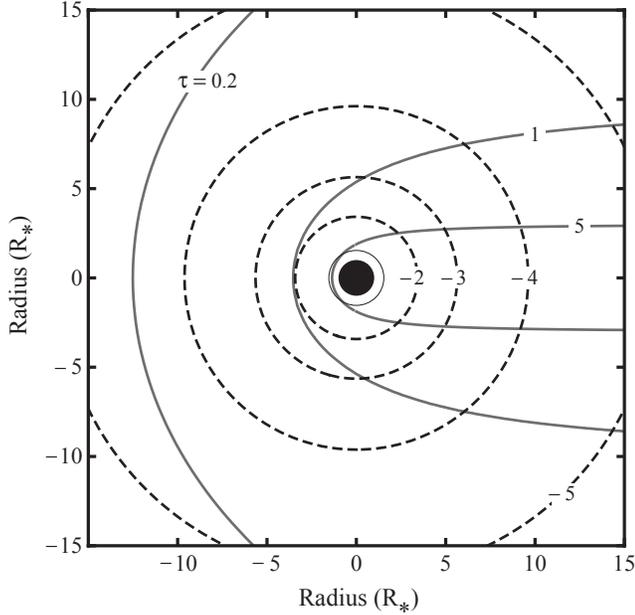}
  \caption{This diagram shows the X-ray emission and absorption
    properties of a model stellar wind. The observer is on the
    left. The black disk denotes the star, and the thin black line
    shows the onset radius for X-ray emission (1.5 $R_*$ in this
    case). Contours of constant X-ray emissivity (proportional to
    density squared) are shown with dashed lines at intervals of an
    order of magnitude in differential emissivity. The absolute scale
    is normalized to the maximum emissivity at the onset radius of
    X-ray emission. Contours of constant continuum optical depth
    calculated for $\tau_* = 3$ are shown with solid lines.}
  \label{fig:james1}
\end{figure}

\begin{figure}
  \begin{center}
  \includegraphics[width=3.3in]{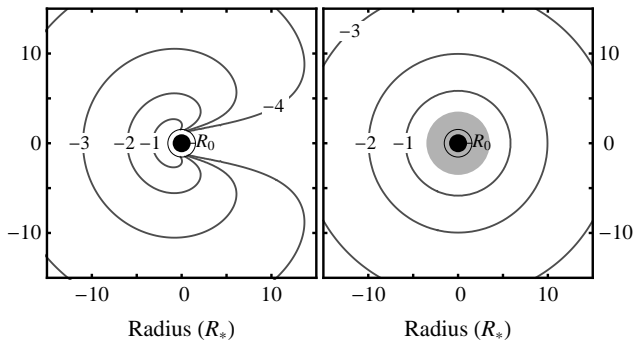}
  \end{center}
  \caption{Left panel is the same as Figure~\ref{fig:james1}, but
    plotting the net emissivity (product of intrinsic emissivity times
    transmission, including occultation by the stellar core) instead
    of the intrinsic emissivity; this net emissivity has been
    renormalized so that the point of maximum net emissivity has a value
    of unity. The right panel shows the same plot in the exospheric
    approximation, where the gray zone is inside the radius of
    (radial) optical depth unity and thus does not contribute to the
    observed X-ray flux.}
  \label{fig:james2}
\end{figure}

The model described in the preceding paragraphs is illustrated in
Figures~\ref{fig:james1} and \ref{fig:james2}. The model emissivity
and transmission are visualized separately in Figure~\ref{fig:james1},
and together in the left panel of Figure~\ref{fig:james2}. The right
panel of Figure~\ref{fig:james2} gives the net transmission for the
exospheric approximation for comparison, which we discuss at more
length below.

Using Equations~(\ref{eq:lx})--(\ref{eq:eta}), we can calculate the transmission
of the wind as a function of $\tau_*$. The transmission is simply the
observed flux (Equation~(\ref{eq:lx})) divided by the unattenuated flux,
which can be found by setting $\tau_* = 0$ in the same equation:
\begin{equation}
  T(\tau_*) \equiv \frac{L_\lambda (\tau_*)}{L_\lambda (0)} = \frac{\int
    dV\, \rho^2\, e^{-\tau}}{\int dV \rho^2}\, .
\end{equation}
To account for the attenuation in the numerator, it is convenient to
define an \emph{angle-averaged transmission} from each radius $r$.
\begin{equation}
  \overline{T}(r, \tau_*) \equiv \frac{1}{2} \int^1_{-\mu_*} d\mu\,
  e^{-\tau_* t(r, \mu)}\, ,
  \label{eq:aat}
\end{equation}
where
\begin{equation}
  \mu_* = \sqrt{1 - \frac{R_*^2}{r^2} }\, 
\end{equation}
gives the $\mu$ coordinate of occultation by the the stellar core.
Some X-rays are obscured by the stellar core even when the wind is
transparent:
\begin{equation}
  \overline{T}_0 (r) \equiv \overline{T} (r, 0) = \frac{1}{2} (1 + \mu_*)\, .
\end{equation}
Integrating over shells at all radii, the net transmission is thus
\begin{equation}
  T(\tau_*) = \frac{\int_{R_0}^\infty\, dr\, r^2\, \rho^2(r)\,
    \overline{T} (r,\tau_*)}{\int_{R_0}^\infty\, dr\, r^2\,
    \rho^2(r)}\, .
\end{equation}
We can further evaluate this expression by substituting the continuity
equation, and by defining the inverse radial coordinate $u \equiv R_*
/ r$:
\begin{equation}
  T(\tau_*) = \frac{\int_0^{u_0}\, du\, w^{-2} (u) \overline{T} (u,
    \tau_*)}{\int_0^{u_0}\, du\, w^{-2} (u)}\, .
  \label{eq:TofTau}
\end{equation}

\begin{figure}
  \includegraphics[width=3.5in]{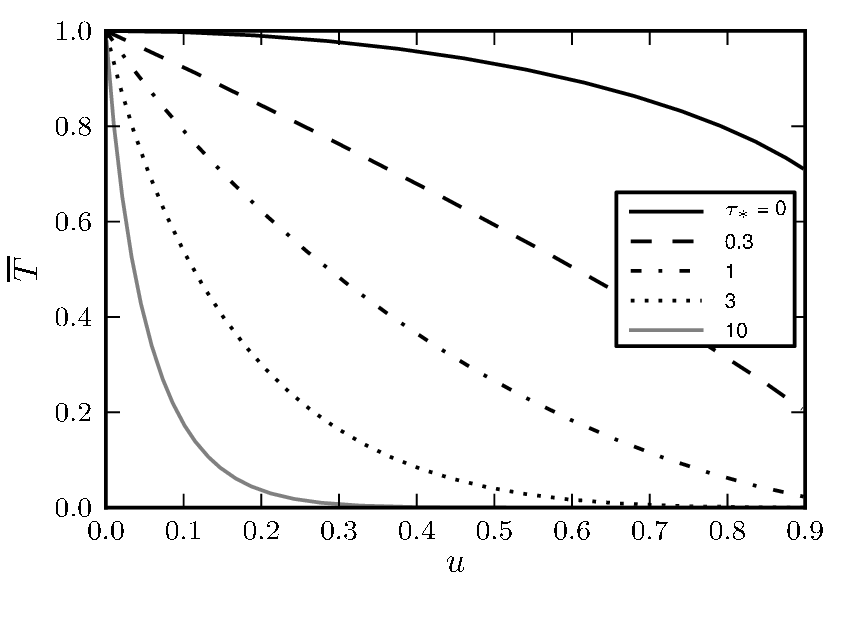}
  \caption{Angle averaged transmission $\overline{T}$ as a function of
    shell inverse radial coordinate $u = R_*/r$, with different curves
    for different characteristic wind optical depths $\tau_*$. The
    equivalent plot in the exospheric approximation would be a step
    function at $u_1$, the inverse radial coordinate of optical depth
    unity.  Note that the transmission is less than unity even for
    $\tau_* = 0$ due to occultation by the stellar core.}
  \label{fig:aat}
\end{figure}

\begin{figure}
  \includegraphics[width=3.5in]{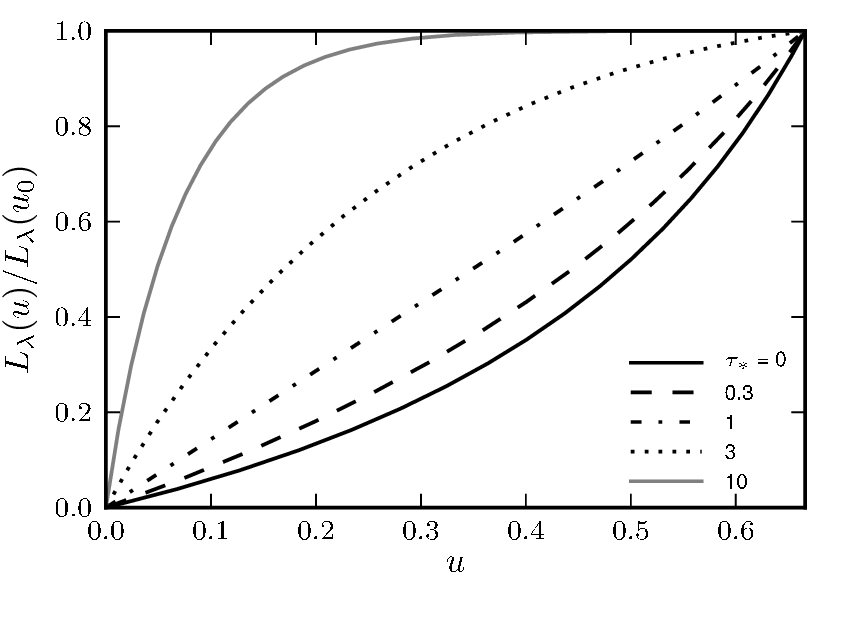}
  \caption{Fraction of X-ray emission originating from outside inverse
    radius $u$, normalized to total X-ray emission integrated to
    $u_0 = 2/3$. For all but the very optically thick case of $\tau_* = 10$,
    the emission comes from a wide range in $u$.}
  \label{fig:cd}
\end{figure}

Figure~\ref{fig:aat} shows the angle-averaged transmission
$\bar{T}(u)$ for selected values of $\tau_*$.
Figure~\ref{fig:cd} shows the following quantity:
\begin{equation}
  \frac{L_\lambda(u, \tau_*)}{L_\lambda (u_0, \tau_*)} =
  \frac{\int^u_0 du^\prime w^{-2} (u^\prime) \overline{T} (u^\prime,
    \tau_*)}{\int^{u_0}_0 du^\prime w^{-2} (u^\prime) \overline{T}
    (u^\prime, \tau_*)}
\end{equation}
This can be thought of as the cumulative distribution of observed
X-ray emission, starting from $u = 0$ ($R = \infty$) and integrating
in to $u = u_0$. Together, Figures~\ref{fig:aat} and \ref{fig:cd} show
the relative importance of transmission and emission as a function of
radius in stellar winds. For the entire range of interest in $\tau_*$,
essentially all of the wind down to $u_0$ contributes to the observed
X-ray flux.

\begin{figure}
  \includegraphics[]{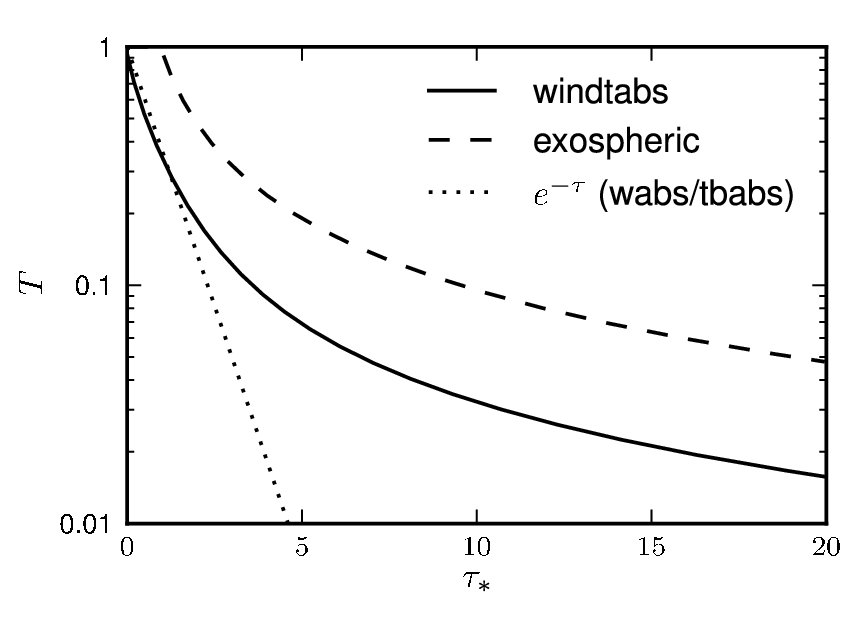}
  \caption{Comparison of transmission of three different models:
    coronal slab ($e^{-\tau}$), exospheric, and more realistic wind model
    (\windtabs). The fixed parameters are $\beta = 1$, $R_0 = 1.5$.}
  \label{fig:transmission}
\end{figure}

Figure~\ref{fig:transmission} shows $T(\tau_*)$ for this model, along
with comparisons to two other absorption prescriptions: a simple
intervening absorber, $T = e^{-\tau}$, appropriate for a coronal slab
model, e.g., as implemented in the XSPEC models \wabs or \tbabs; and
an exospheric approximation \citep[e.g.][]{1999ApJ...520..833O}, where
$T = 0$ below the radius of optical depth unity, and $T = 1$
everywhere above it:
\begin{equation}
  T_{\mathrm{exo}} (\tau_*) = \frac {\int_0^{u_{x}}\, du\,
    w^{-2} (u)}{\int_0^{u_0}\, du\, w^{-2} (u)}\, ,
  \label{eq:Texospheric}
\end{equation}
where $u_{x} \equiv \min (u_1(\tau_*), u_0)$.
The inverse radial coordinate of optical depth unity is given by
evaluating the optical depth integral (Equation~(\ref{eqn:tpz})) along a
radial ray ($p = 0$, $z = r$), with the result
\begin{equation}
  u_1 (\tau_*) = 1 - e^{-1 / \tau_*}\, ,
\end{equation}
for $\beta = 1$.  Note that in
Figures~\ref{fig:aat}-\ref{fig:transmission}, we have used $\beta = 1$
and $R_0 = 1.5 R_*$.  We stress that Equation~(\ref{eq:Texospheric}) is
simply the consequence of using a step function for the angle-averaged
transmission $\overline{T}$ in Equation~(\ref{eq:TofTau}), rather than the
more realistic expression given in Equation~(\ref{eq:aat}).

The transmission of our model falls off much more gradually than
$e^{-\tau}$ (Figure~\ref{fig:transmission}), but it is also more
accurate than the exospheric approximation, especially at moderate
optical depth. The exospheric approximation has the correct asymptotic
behavior for large values of $\tau_*$, but overestimates the
transmission by a fixed factor. The exospheric transmission at large
$\tau_*$ can be brought into agreement with \windtabs by multiplying
the exospheric $\tau_*$ by three.

It would be possible to generalize the radiation transport
model described in this section to include porosity by introducing an
appropriate definition of effective opacity, as in \citet{OFH06} or
\citet{OC06}. However, a specific implementation of this and
discussion of its consequences is left to a future study.

\section{Opacity model}
\label{sec:opacity}

To determine the wind transmission as a function of wavelength, we
need to account for the wavelength dependence of the optical depth
through the opacity, which we write as
\begin{equation}
  \tau_*(\lambda) = \kappa(\lambda) \Sigma_*\, ,
\end{equation}
where
\begin{equation}
  \Sigma_* \equiv \frac{\dot{M}}{4\pi R_* v_\infty}
  \label{eq:sigmastar}
\end{equation}
is the {\it characteristic mass column density} of the wind (in
$\mathrm{g}\, \mathrm{cm}^{-2}$).

The continuum opacity of a stellar wind in the X-ray band can be
calculated by summing the contributions of each constituent
species. Thus, we must know the atomic cross-sections, ionization
fractions, and elemental abundances. Of these three, the cross
sections are known to sufficient accuracy \citep[e.g.][]{VY95};
the ionization balance may contribute some uncertainty in the
calculation of the opacity, but is usually not the dominant source of
error; and uncertainties in the elemental abundances are typically the
most important.

The opacity due to photoionization of a given shell of any individual
species scales approximately as $\kappa_i (\lambda) \propto
\lambda^{3}$ above the threshold energy $E_\mathrm{th}$ of the
shell. Because multiple species are usually important for the X-ray
opacity of astrophysical gas, the run of opacity with wavelength has a
characteristic sawtooth shape, with individual teeth at the ionization
threshold energies of dominant ionization stages of abundant elements.  

O star winds are photoionized, with H and He fully stripped (although
He may recombine in some dense winds; see below), and most other
elements mainly in charge states +3 and +4. Thus, the opacity of
stellar winds in the range 1 \AA\ $< \lambda <$ 40 \AA\ is dominated by
K-shell absorption in C, N, and O, since they are the most abundant
elements. Significant contributions from K-shell absorption in Ne, Mg,
and Si are also present, as well as Fe L-shell absorption.

The opacity of adjacent ionization stages of the same element are
usually comparable, with the exception that the photoionization
threshold energy is shifted.  The effect of a moderate shift in
ionization balance on the opacity is relatively minor, since none of
the strong X-ray emission lines observed from O star winds falls
between the threshold energies of adjacent stages of abundant ions.
However, the difference in threshold energies, and hence the broadband
absorption, between neutral and wind material is significant.

Therefore, while it is important to use an appropriate model for the
wind ionization, it is sufficient for many applications
to use a single approximate ionization balance to model all O star
winds. This is true even though the ionization balance can vary to
some extent with radius, and also is different in different stars.

If it is desired to model the opacity of a particular star, it is
possible to construct a detailed model opacity for a stellar wind by
using the output of a radiative transfer model such as CMFGEN
\citep[J. Zsarg\'{o} et al., in preparation]{Hel93,
  1998ApJ...496..407H, Cohen2010}.

\begin{figure}
  \includegraphics[width=3.5in]{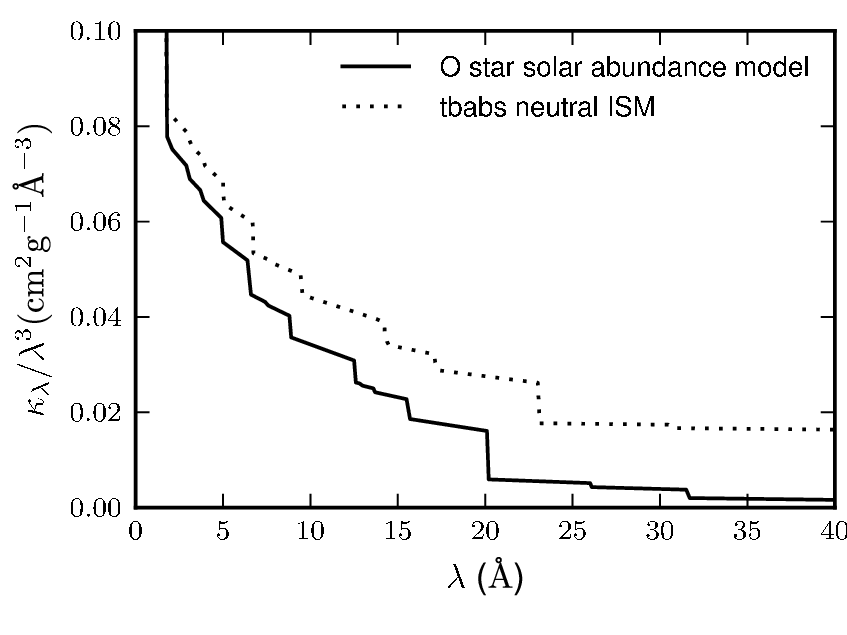}
  \caption{Comparison of a neutral interstellar absorption model
    (including dust grains), \tbabs \citep{WAM2000}; and an O star
    wind model with an assumed simple ionization structure. The
    opacity $\kappa(\lambda)$ is scaled with $\lambda^3$ to allow a
    better comparison of the two models at short wavelengths. Both
    models use solar abundances \citep{2009ARA&A..47..481A}. Note that
    even below the O$^{3+}$ edge near 20 \AA, the wind opacity model
    is still about 40\% lower than the ISM model, mainly due to the
    ionization of H and He in the wind model.}
  \label{fig:opacityISM}
\end{figure}

\begin{figure}
  \includegraphics[width=3.5in]{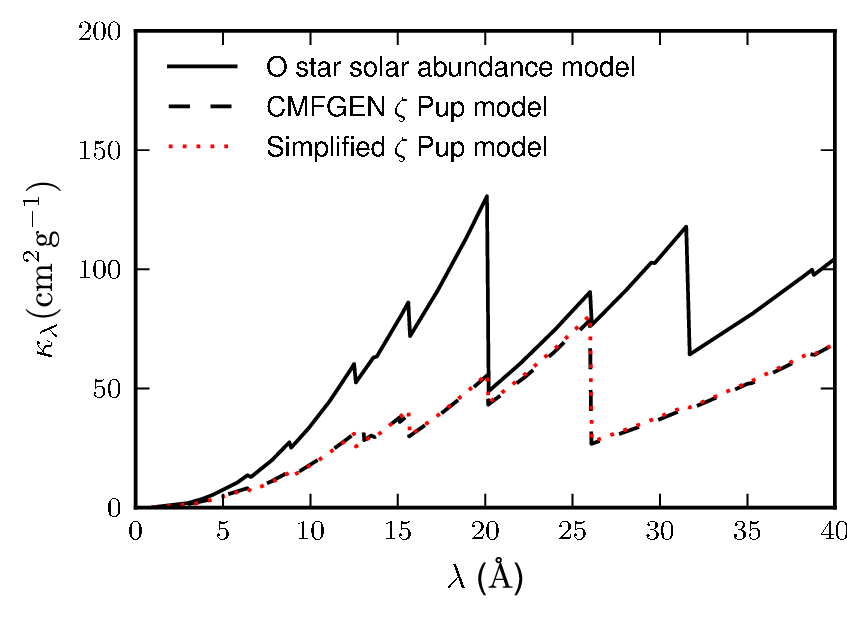}
  \caption{Comparison of three different model O star wind opacities:
    the solar abundance O star wind model (with assumed simple
    ionization structure) shown in Figure~\ref{fig:opacityISM}
    (solid); a CMFGEN model appropriate for \zp (dashed) using the
    abundances of J.-C. Bouret et al.\ (in preparation); and a
    simplified version of this CMFGEN model (dotted; red in the
    electronic version) where the ionization balance is the same as in
    the solar abundance model.  Note that the Bouret et
    al.\ abundances for \zp are subsolar as well as having altered CNO
    abundance ratios.}
  \label{fig:opacity}
\end{figure}

To illustrate the importance of various assumptions in opacity
modeling, Figures~\ref{fig:opacityISM} and \ref{fig:opacity} compare
several model opacities. Figure~\ref{fig:opacityISM} compares neutral
interstellar medium opacity and a simple model for the opacity of a
stellar wind. Both assume solar abundances
\citep{2009ARA&A..47..481A}. In the wind model we assume an ionization
balance with hydrogen and helium fully stripped, oxygen and nitrogen
in the +3 charge state, and all other elements in the +4 charge
state. This is a good approximation to the ionization balance in a
typical O star wind, which is set by photoionization from the
photospheric UV field. The model wind opacity is much lower than the
model ISM opacity, especially at long wavelengths, mainly due to the
ionization of hydrogen and helium. The shift in ionization threshold
energies due to the presence of more highly ionized species is also
clear.

Figure~\ref{fig:opacity} shows three stellar wind model opacities, and
thereby illustrates the relative importance of two effects on the
opacity: the elemental abundances, and the ionization balance. The
solid line gives the same solar abundance wind model described in the
previous paragraph, while the dashed and dotted lines give models
particular to \zp, using non-solar abundances specific to the star
derived from detailed CMFGEN modeling (J.-C. Bouret et al., in
preparation).  The dashed line gives the actual opacity at $\sim 2
R_*$ from this CMFGEN model, while the dotted line uses the simplified
ionization structure of the solar abundance wind opacity model. The
fact that the realistic CMFGEN \zp model and the simplified version
are so similar indicates that a simple ionization balance is typically
adequate to describe wind opacity in many cases, as long as the
dominant ionization stages are relatively accurate. On the other hand,
the difference between the \zp models and the solar abundance model
shows that the opacity model depends strongly on the abundances of the
most common elements other than H and He (typically C, N, and O).
Note that the Bouret et al.\ abundances for \zp are both non-solar in
the ratio of CNO and {\it also} have sub-solar metallicity; it is the
sub-solar metallicity that accounts for the lower opacity of the \zp
models at short wavelengths compared to the solar abundance model.

We have made one important simplification in our modeling: in
Equation~(\ref{eq:tautotaustar}), and throughout this section, we have
assumed that the opacity is independent of radius. As shown in
Figure~\ref{fig:opacity}, moderate changes in wind ionization do not
strongly affect the opacity, so in most cases this is a justified
approximation. The important exception is the ionization of helium; in
sufficiently dense winds, He$^{++}$ may recombine to He$^+$ in the
outer part of the wind, which greatly increases the opacity,
especially at long wavelengths \citep{Pauldrach87, Hel93}. As long as
the change in ionization occurs sufficiently far out in the wind,
geometrical effects as described in Section~\ref{sec:radtran} are not
important, and the absorption due to He$^+$ can be treated as an
additional slab between the X-ray emitting regions and the observer,
i.e., using $T = e^{-\tau}$.

\section{Model Implementation}

The numerical evaluation of Equation~(\ref{eq:TofTau}) is not prohibitively
expensive, but it is typically not fast enough to allow its use in an
automated spectral fitting routine, such as that in XSPEC12
\citep{1996ASPC..101...17A} or ISIS \citep{2000ASPC..216..591H}. It is
thus preferable to compute the transmission on a grid in $\tau_*$ for
a given set of wind parameters.

Given a tabulation of the model wind opacity, as described in
Section~\ref{sec:opacity}, in addition to the tabulation of $T(\tau_*)$,
one may then calculate the transmission as a function of wavelength,
$T(\lambda)$, with \emph{only one free parameter}, the characteristic
wind mass column density $\Sigma_*$. This parameter is analogous to
the neutral hydrogen column density in a slab absorption model such as
the XSPEC \wabs or \tbabs ISM absorption models (which are also
sometimes used to approximate wind absorption).

We have implemented this as a local model for XSPEC 12\footnote{Source
  code is freely available under the General Public License, and it
  may be obtained along with installation instructions at
  \url{http://heasarc.gsfc.nasa.gov/docs/xanadu/ xspec/models/windprof.html}}.
A user can calculate $T(\tau_*)$ for a given set of parameters
(i.e., $\beta$, $R_0$), with the results stored in a FITS table. These
implicit model parameters may be varied by computing additional tables
of $T(\tau_*)$ for each set of parameter values. However, the
absorption model is not very sensitive to these parameters over the
range typically inferred for winds of massive stars. The calculation
of $T(\tau_*)$ is controlled by a simple python script, and
computation of a table for a given set of parameters can be
accomplished in several seconds on a modern workstation.  The model
opacity must also be supplied as a FITS file; different model
opacities may be swapped in at run time. When \windtabs is used in a
spectral fitting program, model transmission is calculated as a
function of energy or wavelength using the supplied FITS tables and
the one free model parameter, $\Sigma_*$.

The elemental abundances, which enter into the transmission through
their effect on the opacity, cannot be varied as fit parameters in our
model. This is a choice we have made in the model implementation, both
for computational ease and simplicity of user interface, and because
there is not enough information in X-ray spectra to constrain
elemental abundances through modeling of \emph{absorption} of X-ray
emission alone. Abundances should be inferred by other means,
e.g., from global fits to UV and optical spectra, or from fits to X-ray
\emph{emission line strengths}. These abundances can be used to
compute a new opacity table for a given star.

\begin{figure}
  \includegraphics[width=3.5in]{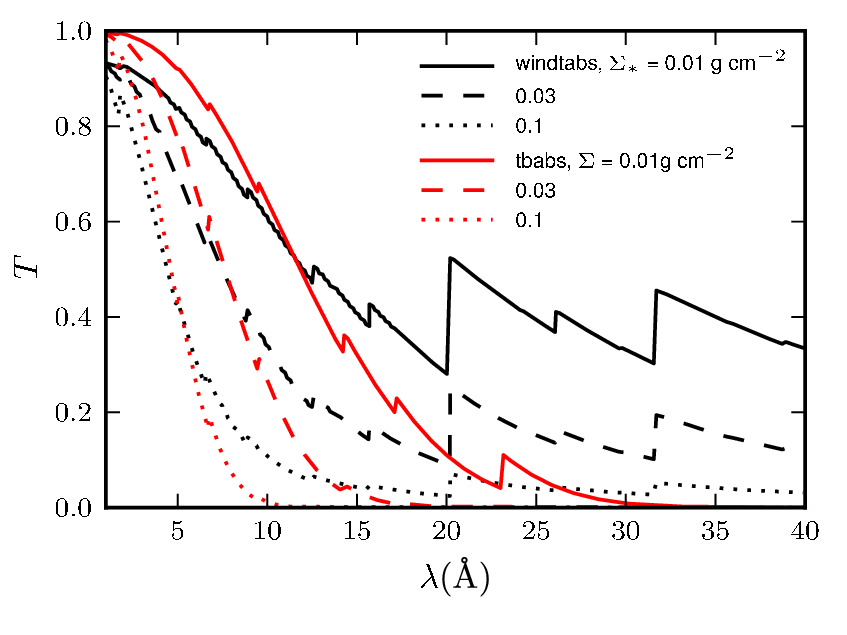}
  \caption{Transmission as a function of wavelength for ionized wind
    absorption model (\windtabs, black) and for neutral slab
    absorption (\tbabs, gray; red in the electronic version). Three
    values of absorbing column are given; for \windtabs, the degree of
    absorption is specified by the characteristic mass column density
    $\Sigma_*$, while for \tbabs it is given simply by the mass column
    $\Sigma$.}
  \label{fig:transmissionWavelength}
\end{figure}

Figure~\ref{fig:transmissionWavelength} gives the model transmission
for \windtabs using three different values of $\Sigma_*$, and using
our standard O star wind opacity model. For comparison, this figure
also shows the transmission for the neutral absorption model \tbabs
for comparable mass column densities $\Sigma$. Note that $\Sigma$
refers simply to a slab mass column density, while $\Sigma_*$ refers
to a {\it characteristic} mass column density in the context of a
stellar wind (see Equation~(\ref{eq:sigmastar})).

\section{Discussion}

\subsection{Advantages of \windtabs over exospheric or exponential attenuation}

The exact method we have presented here for modeling the X-ray
radiative transfer for a distributed source embedded within a stellar
wind is crucial for analyzing the X-ray emission observed from O
stars.  Compared to the exponential, neutral slab absorption (excess
over ISM) model that is usually employed, the \windtabs model
accurately reproduces the much more gradual decrease in transmission
with increasing wind column density and opacity. This can be seen in
Figure~\ref{fig:transmissionWavelength}, where the exponential
transmission model shows an unrealistically sudden decrease in
transmission when the wind becomes optically thick.

Because the opacity of the bulk wind is a relatively strong function
of wavelength, the inaccuracy of the exponential transmission model
will lead to errors in the broadband spectral energy distribution of a
model applied to individual O stars, leading to misinterpretations of
the associated emission model components.  This appears to be the case
in the study of \cxo grating spectra in
\citet{2007MNRAS.382.1124Z}, which invokes excess exponential, neutral
ISM absorption to account for the assumed wind attenuation. Most
likely because the exponential transmission radiation transport model,
and also the neutral ISM opacity model, significantly overestimates
the degree of attenuation, these authors find absorption beyond the
ISM column for only one star, even though some additional wind
attenuation is expected for most of the stars in their sample.
Additionally, although the authors do not comment on it, their
determination of elemental abundances for each of the stars shows a
consistent trend of abundance correlated with the wavelength of the
dominant line or lines from each element.  Such an effect would be
expected if the wavelength-dependent wind attenuation is not
accurately accounted for.

It has long been noted that the exponential attenuation treatment is
not well suited to modeling OB star X-rays. \citet{Hel93} used an
exact treatment in their modeling of $\zeta$ Pup, discarding the
possibility of a coronal model on the basis of the strong soft
observed X-ray flux of \zp, together with its relatively high
mass-loss rate. \citet{Cohenel96} showed that an exospheric treatment,
rather than an exponential treatment, is important for understanding
the observed EUV and soft X-ray emission from the early B giant,
$\epsilon$ CMa. The exospheric approximation was used by
\citet{1999ApJ...520..833O} to model the effect of wind attenuation of
X-rays in order to explain the observed $L_x/L_{\mathrm{bol}} \sim 10^{-7}$
relationship and its breakdown in the early B spectral range where hot
star winds become optically thin to X-rays.  An exospheric treatment
was also used by \citet{2001A&A...373.1009O} to analyze the
variability of X-rays from optically thick WR winds.  And the
exospheric framework forms the basis for the ``optical depth unity''
relationship for X-rays in O stars, where the
forbidden-to-intercombination line ratios of helium-like ions are
claimed to imply formation radii that track the optical depth unity
radius as a function of X-ray wavelength \citep{WC07}.

However, the exospheric treatment underestimates the attenuation of
the wind. If data need to be analyzed with a high degree of accuracy,
then a more realistic treatment of X-ray radiation transport through
the wind must be used; one that takes the inherently non-spherically
symmetric nature of the problem into account.  This is especially true
when the location of the X-ray emission emerging from the wind is
important, as with the interpretation of $f/i$ ratios.  As
Figure~\ref{fig:james2} shows, the relative contribution of the
emergent X-ray flux from different wind regions is incorrectly
estimated in the exospheric approach, and as Figure~\ref{fig:cd}
shows, the emergent flux has significant contributions from a wide
range of radii. In fact, a number of previous works have implemented
accurate X-ray radiative transfer prescriptions \citep[e.g.,][]{PHL01,
  OC01, OFH06}; however, none of these works provide an implementation
that is available for use in other contexts.

The \windtabs model we have introduced here is not only more accurate
in terms of the radiation transport than slab or exopheric treatments,
but it has two additional advantages that recommend its adoption for
routine X-ray data analysis and modeling of O stars.  First, it is
easy to use and has only a single free parameter, the characteristic
mass column density $\Sigma_*$, from which a mass-loss rate can be
readily extracted.  And second, it incorporates a default wind opacity
model that is significantly different from, and much more accurate
than, the neutral ISM opacity models that are usually used.
Additionally, alternate user-calculated opacity models are easy to
incorporate.

\subsection{Implications for trends in X-ray hardness with spectral type}

One application of \windtabs to the interpretation of X-ray spectral
data is for the analysis of the X-ray spectral hardness trend
versus optical spectral subtype recently noted in \cxo grating spectra by
\citet{WNW09}. They report ``the progressive weakening of the higher
ionization relative to the lower ionization X-ray lines with advancing
spectral type, and the similarly decreasing intensity ratios of the
H-like to He-like lines of the $\alpha$ ions.'' The correlation of the
overall X-ray spectral hardness with spectral subtype described by
Walborn et al.\ appear to be a direct result of wavelength dependent
absorption effects. A detailed analysis is in preparation, but here we
show a single suite of models in which the only variable parameter is
the characteristic wind column density, $\Sigma_*$. A single emission
model, combined with \windtabs attenuation, reproduces the observed
broadband trend quite well.

\begin{figure*}
  \includegraphics[angle=-90]{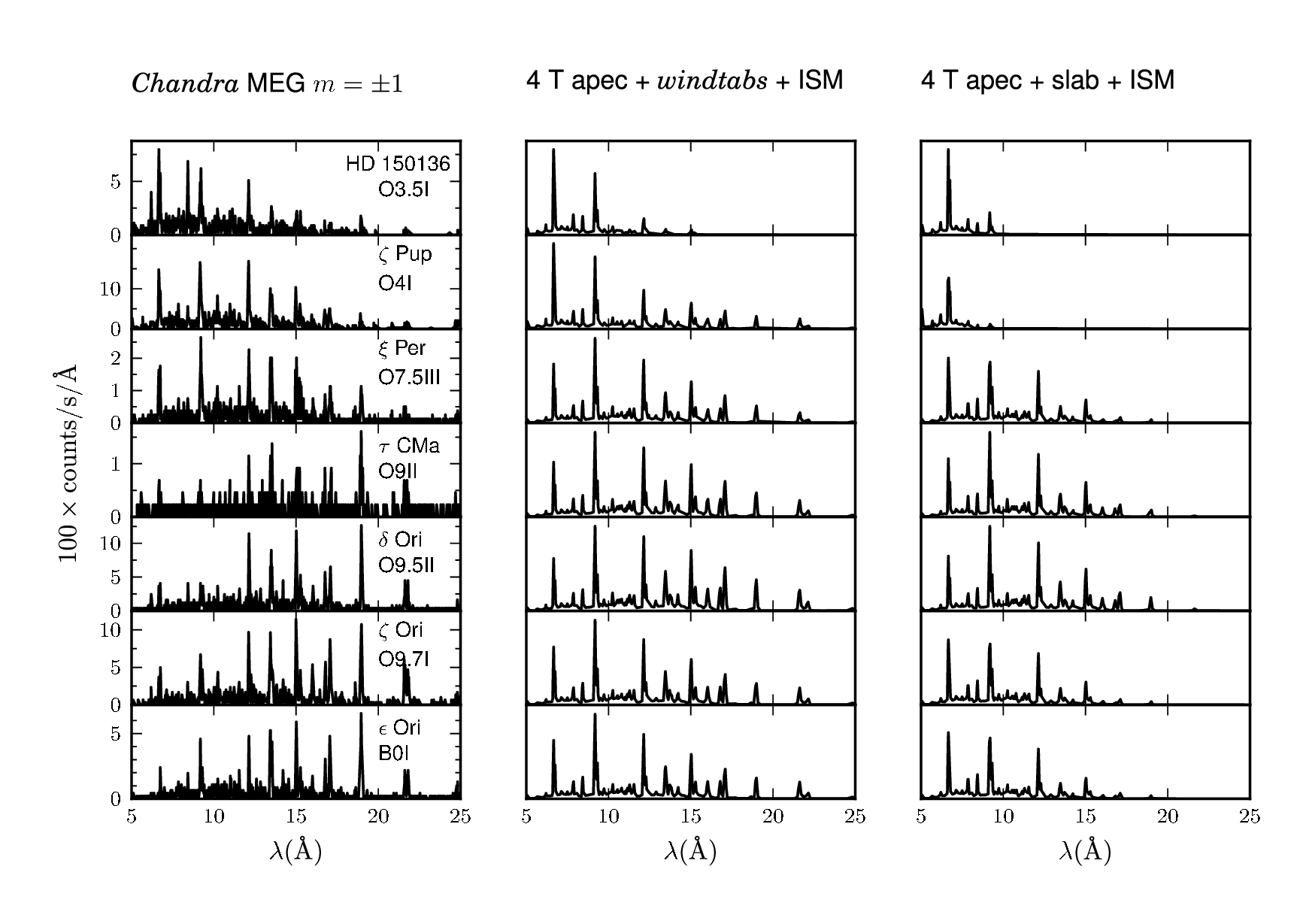}
  \caption{Left column: a sequence of \cxo spectra of O
    giants and supergiants from \citet{WNW09}; middle column:
    multi-temperature thermal emission model with \windtabs wind
    absorption model; right column: same model as middle column, but
    with \tbabs neutral slab absorption model.}
  \label{fig:spectralTypeTrend}
\end{figure*}

Figure~\ref{fig:spectralTypeTrend} shows X-ray grating spectra for
seven O giants and supergiants in the left-hand column, with the
earliest spectral subtype (O3.5) on the top, and the latest (B0) on
the bottom, as in \citet{WNW09}.  The later spectral subtypes clearly
have more soft X-ray emission, although the earlier subtypes still
have non-negligible long-wavelength ($\lambda \gtrsim 15$ \AA)
emission.  In the middle column we show a four-temperature \apec
\citep{SBLR01} thermal equilibrium emission model. We have chosen $kT$
= 0.1, 0.2, 0.4, and 0.6 keV; the first three components having equal
emission measures and the hottest one having half the emission measure
of the others.  The same \apec model is used for all seven stars (with
variable overall normalization) and is multiplied by a \windtabs model
and a \tbabs model (for neutral ISM attenuation).  The column density
of the \tbabs model is fixed at the interstellar value taken from
\citet{1994ApJS...94..127F}, with the exception of HD 150136, for
which we inferred the ISM column density from E(B-V)
\citep{2004ApJS..151..103M, 2006A&A...457..637M, 2003A&A...408..581V}.
The characteristic mass column density $\Sigma_*$ in \windtabs is
fixed at a value computed from the ``cooking formula'' theoretical
mass-loss rate computed by \citet{2001A&A...369..574V}, using the
measured terminal velocity of \citet{Haser95PHD} and the modeled radii
of \citet{2005A&A...436.1049M}.  The standard solar abundance wind
opacity model (solid line in Figure~\ref{fig:opacity}) was used in
\windtabs, and the \apec model abundances were set to solar.  There
are no free parameters in these models, and the temperature
distribution has not even been significantly optimized to match the
data. The adopted parameters are listed in
Table~\ref{tab:stellarparameters}.

\begin{deluxetable}{ccccc}
\tablecaption{Adopted Stellar Parameters\label{tab:stellarparameters}}
\tablehead{
\colhead{Star} & 
\colhead{Spectral Type\tablenotemark{a}} &  
\colhead{$N_{\mathrm H}$\tablenotemark{b}} &
\colhead{$\Sigma_*$\tablenotemark{c}} &
\colhead{$N_*$\tablenotemark{d}}
}
\startdata
HD 150136 & O3.5 If* + O6V & 0.76 & 0.073 & 3.6 \\
$\zeta$ Pup & O4 I(n)f & 0.01 & 0.160 & 7.2 \\
$\xi$ Per & O7.5 III(n)((f)) & 0.115 & 0.017 & 0.76 \\
$\tau$ CMa & O9 II & 0.056 & 0.013 & 0.58 \\
$\delta$ Ori & O9.5 II + B0.5 III& 0.015 & 0.011 & 0.49 \\
$\zeta$ Ori & O9.7 Ib & 0.03 & 0.019 & 0.85 \\
$\epsilon$ Ori & B0 Ia & 0.03 & 0.020 & 0.89 
\enddata
\tablenotetext{a}{Spectral types adopted by \citet{WNW09}.}
\tablenotetext{b}{Interstellar medium column density ($10^{22}\, \mathrm{cm}^{-2}$).}
\tablenotetext{c}{Characteristic wind mass column density (g cm$^{-2}$).}
\tablenotetext{d}{Characteristic wind number column density equivalent to $\Sigma_*$ ($10^{22}\, \mathrm{cm}^{-2}$).}
\end{deluxetable}

As the middle column of Figure~\ref{fig:spectralTypeTrend} shows, the
simple, universal emission model reproduces the broadband trend very
well.  Trends in individual line ratios generally cannot be reproduced
only by accounting for the varying attenuation, as pointed out by
\citet{WNW09}, but note that the \ion{Ne}{9} (13.5 \AA) to \ion{Ne}{10}
(12.1 \AA) ratio does indeed vary due only to differential attenuation
among the earliest spectral subtypes. The right-hand column in
Figure~\ref{fig:spectralTypeTrend} shows the same emission and ISM
attenuation models as in the middle column, but with excess
exponential (neutral ISM) attenuation accounting for the wind
absorption, again according to the wind column densities expected from
the adopted mass-loss rates, radii, and terminal velocities.  The
exponential attenuation trend seen in the right-hand column is too
strong for the earliest spectral subtypes and too weak for the latest
ones, where the different ISM column densities actually dominate the
trend.

The contrast between the \windtabs and exponential models is quite
stark, and indicates that the more realistic models should generally
be used when analyzing X-ray spectra, both high-resolution and
broadband. It is also impressive how much of the observed spectral
hardness trend is explained by wind attenuation, in the context of a
realistic model.  Not only do quantitative analyses of the suggested
line ratio trends have to be evaluated, but a global spectral modeling
that allows for both emission temperature variations and wind
attenuation variations should be undertaken in order to disentangle
the relative contributions of trends in emission and absorption to the
overall, observed trend in the spectral energy distributions.

\section{Conclusions}

We have presented an exact solution to the radiation transport of X-rays
through a spherically symmetric, partially optically thick O star wind,
and shown that it differs significantly from the commonly used slab
absorption and exospheric models.  Specifically, the transmission falls
off much more gradually as a function of fiducial optical depth in the
\windtabs model as compared to the exponential model, leading to more
accurate assessments of wind column densities and mass-loss rates from
fitting X-ray spectra.  As one example of the utility of \windtabs, we
have shown that when this more accurate model is employed, differential
wind absorption can explain most of the observed trend in OB star X-ray
spectral hardness with spectral subtype, and even may explain some of
the line ratio trend.

The \windtabs model has been implemented as a custom model in XSPEC,
and is as easy to use as the various ISM absorption models, having
only one free parameter.  In addition to the significantly improved
accuracy of the radiation transport, \windtabs has several other
advantages.  It incorporates a default opacity model much more
appropriate to stellar winds than the neutral element opacity model
used in ISM attenuation codes.  Users can easily substitute their own
custom-computed opacity models.  And the fitted mass column density
parameter for \windtabs allows for the user to extract a mass-loss
rate directly from their fitting of X-ray spectra of OB stars.

\acknowledgements

M.A.L. is supported by an appointment to the NASA Postdoctoral Program at
Goddard Space Flight Center, administered by Oak Ridge Associated
Universities through a contract with NASA.  D.H.C. acknowledges support
from \cxo awards AR7-8002X and GO0-11002B to Swarthmore College,
and D.H.C., E.M.M., and J.P.M. thank the Provost's Office at Swarthmore College
for support via Eugene M. Lang and Surdna Summer Research
Fellowships. J.Z. and D.J.H. were supported by STScI grant HST-AR-10693.02
and by SAO grants TM6-7003X and GO0-11002A.
\bibliography{xray-ostar}

\end{document}